\documentclass[11pt]{article}
\usepackage{amsmath}

\numberwithin{equation}{section}

\begin{document}
\baselineskip=15pt
\begin{titlepage}
\begin{flushright}
{\small KYUSHU-HET-78}\\[-1mm] hep-ph/0501211
\end{flushright}
\begin{center}
\vspace*{10mm}

{\Large\bf%
Asymmetry and Minimality of Quark Mass Matrices%
}\vspace*{8mm}

Nobuhiro Uekusa, Atsushi Watanabe, and Koichi Yoshioka
\vspace*{2mm}

{\it Department of Physics, Kyushu University, Fukuoka 812-8581, Japan}
\vspace*{5mm}

{\small (January, 2005)}
\end{center}
\vspace*{5mm}

\begin{abstract}\noindent%
We systematically investigate general forms of mass matrices for
three-generation up and down quarks, including asymmetrical ones in
generation space. Viable zero matrix elements are explored which are
compatible with the current observation of masses and mixing angles,
and also with the recent measurement of CP violation in the $B$-meson
system. The simplest form with the maximal number of vanishing matrix
elements is found to be almost consistent with the experimental data,
but has a scratch that one of the mixing angle is slightly large. At
the next-to-minimal level, it is found with a help of leptonic
generation mixing that only six patterns of mass matrices well
describe the experimental data. These sets of mass textures predict
all the properties of quarks, including the CP violation, as well as
the large (charged) lepton mixing, which may be appropriate for the
atmospheric neutrino in grand unification scheme.
\end{abstract}
\end{titlepage}

\section{Introduction}

It is certain that one of the most important issues confronting the
standard model is the generation structure of quarks and
leptons. Various neutrino oscillation experiments have recently been
bringing out the generation structure of the leptonic sector. The
Super-Kamiokande experiment has established the neutrino oscillation
in the atmospheric neutrinos with nearly maximal
mixture~\cite{SKatm}. As for the solar neutrino problem, the
Mikheyev-Smirnov-Wolfenstein solution~\cite{MSW} is strongly suggested
by the recent experimental results~\cite{solar-exp} if there exists
the neutrino flavor mixing between the first and second
generations~\cite{analyses}. On the other hand, the other mixing (the
1-3 mixing) has been found to be rather small~\cite{CHOOZ} similarly
to the quark sector. While these experimental progresses have been
giving us a new perspective beyond the standard model, it seems that
we are far from a fundamental understanding of the origin of fermion
masses and mixing angles.

A promising approach to the issue of the generations is to assume that
some of Yukawa matrix elements are vanishing. An immediate and
important consequence of this approach is to reduce the number of free
parameters in the theory and to lead to relations among the fermion
masses and mixing angles. Moreover that would provide a clue to find
symmetry principles or dynamical mechanisms behind the Yukawa sectors,
which are interpreted as remnants of fundamental theory in high-energy
regime. Most of the previous work along this direction, including the
systematic analysis by Ramond, Roberts and Ross~\cite{RRR}, assumed
that the matrices of Yukawa couplings are symmetric about the
generation indices (hermitian matrices).\footnote{Systematic
studies and classifications of fermion mass matrices and mixing have
also been performed in other approaches~\cite{others}.} However, in
the context of the standard model and even in grand unified theory, it
is not necessarily required that Yukawa matrices are symmetric. In
fact, the present experimental data indicate that the (large) leptonic
mixing mentioned above is quite different from the (small) quark
mixing. It is interesting that this asymmetrical observation is known
to be compatible with quark-lepton grand unification if the fermion
mass matrices take asymmetrical forms in generation
space~\cite{lopsided}. There are also some classes of 
non-hermitian {\it ansatze} for quark mass matrices~\cite{non}, which
are consistent with the data and cannot be transformed to the
solutions obtained in Ref.~\cite{RRR}. It is therefore worthwhile to
do systematical examination and to complete the classification of
viable asymmetric forms of fermion mass matrices.

In this paper, we investigate phenomenologically viable mass matrices
of up and down quarks, assuming that the Yukawa couplings generally
take asymmetric forms in generation space. In particular, we look for
as simple forms as possible, that is, mass matrices with the maximal
number of vanishing elements. Vanishing matrix elements are expected
to be deeply connected with underlying physics, such as flavor
symmetries, in more fundamental theory to shed some lights on
constructing realistic models of quarks and leptons. Note that our
treatment is general and includes symmetric mass matrices as limited
cases.

This paper is organized as follows. In Section~2, we describe the
Yukawa sectors of up and down quarks in the standard model and
introduce the parameterization needed in later discussion. Our
analysis does not depend on any details of the Higgs field profile,
and therefore can be straightforwardly applied to other cases such as
grand unified theory and supersymmetric models. Sections~3 and 4 are
devoted to analyzing which forms of matrices (vanishing matrix
elements) are compatible with the current experimental data. It is
found that the minimal (Section~3) and next-to-minimal (Section~4)
cases contain only a few types of mass matrices phenomenologically
viable. We summarize our results in Section~5.

\section{Formulation}

In this section, we briefly review the formulation of the up and down
quark Yukawa sectors in the standard 
model. The $SU(3)\times SU(2)\times U(1)$ gauge-invariant Yukawa
interactions are given by
\begin{equation}
  -{\cal L}_Y \,=\, \bar Q_i (Y_u)_{ij} u_R{}_j H^* 
  +\bar Q_i (Y_d)_{ij} d_R{}_j H +\textrm{h.c.},
\end{equation}
where $Q_i$ denote the $SU(2)$ doublets of left-handed quarks, and
$u_R,d_R$ are the right-handed up- and down-type quarks, 
respectively. The Yukawa couplings $Y_u$ and $Y_d$ are 3$\times$3
matrices ($i,j$ the generation indices), and $H$ is the $SU(2)$
doublet Higgs field. After the electroweak symmetry breaking, these
Yukawa interactions lead to the following quark mass terms:
\begin{gather}
  -{\cal L}_m \,=\, \bar u_L{}_i (M_u)_{ij} u_R{}_j 
  +\bar d_L{}_i (M_d)_{ij} d_R{}_j +\textrm{h.c.}, \\[1mm]
  \quad (M_u)_{ij} = (Y_u)_{ij} v, \qquad (M_d)_{ij} = (Y_d)_{ij} v, 
  \nonumber
\end{gather}
where $v$ is a vacuum expectation value of the neutral component of
the Higgs field $H$. If the model is supersymmetrized, the Yukawa
terms are described by superpotential in terms of quark and Higgs
superfields. Only one difference is that in the supersymmetric case
two types of Higgses must be introduced to have gauge-invariant Yukawa
terms. This procedure does not bring any modification to the structure
of Yukawa couplings $Y_{u,d}$, and therefore the following analysis is
straightforwardly extended to supersymmetric models and also to other
scenarios.

The generation mixing is physically described by the
Cabibbo-Kobayashi-Maskawa (CKM) matrix which consists of two unitary
matrices
\begin{equation}
  V_{\rm CKM} \,=\, V_{uL}^\dagger V_{dL}.
\end{equation}
These unitary matrices diagonalize the mass matrices $M_u$ and $M_d$;
\begin{equation}
  M_u \,=\, V_{uL} M_u^D V_{uR}^\dagger, \qquad
  M_d \,=\, V_{dL} M_d^D V_{dR}^\dagger.  
\end{equation}
The diagonal elements of $M_u^D$ and $M_d^D$ correspond to the
experimentally observed mass eigenvalues.

With phase degrees of freedom of the six quark fields, the number of
observable parameters in the CKM matrix is reduced to four (the
overall phase rotation is physically irrelevant). However, it is
important to distinguish the contributions of $V_{uL}$ and $V_{dL}$
from a viewpoint of pursuing clues to find more fundamental theory of
quarks and lepton such as grand unification and flavor symmetry. A
generic $3\times 3$ unitary matrix $U$ has 9 free parameters and
can be parameterized as
\begin{equation}
  U \,=\, \Phi O_1\Phi'O_2O_3\Phi''.
  \label{U}
\end{equation}
The matrices $O_i$ ($i=1,2,3$) represent the rotations in the index
space around the $i$-th axis
\begin{equation}
O_1 = 
\begin{pmatrix}
  1 & \!0 & \!0 \\
  0 & \!\cos\theta_1 & \!\sin\theta_1 \\
  0 & \!-\sin\theta_1 & \!\cos\theta_1
\end{pmatrix}\!,\quad
O_2 =
\begin{pmatrix}
  \cos\theta_2 & \!0 & \!\sin\theta_2 \\
  0 & \!1 & \!0 \\
  -\sin\theta_2 & \!0 & \!\cos\theta_2
\end{pmatrix}\!, \quad
O_3 =
\begin{pmatrix}
  \cos\theta_3 & \!\sin\theta_3 & \!0 \\
  -\sin\theta_3 & \!\cos\theta_3 & \!0 \\
  0 & \!0 & \!1
\end{pmatrix}.
\end{equation}
The diagonal phase matrices $\Phi$'s are given by 
$\Phi=\mathrm{diag.}(e^{i\phi},e^{i\varphi},1)$, 
$\Phi'=\mathrm{diag.}(1,1,e^{i\omega})$ and 
$\Phi''=\mathrm{diag.}(e^{ip},e^{iq},e^{ir})$. There generally exists
three rotation angles and six complex phases. When applied to the
above quark mixing matrices $V_{uL}$ and $V_{dL}$, the phases factors
in $\Phi''$ are always unphysical degrees of freedom since they can be
absorbed by field redefinitions of the quark mass eigenstates. It will
also be found that phase matrices $\Phi'$'s do not appear throughout
this work (except for only a few examples discussed at the beginning
of Section~4) because we will consider $3\times 3$ matrices with
non-vanishing determinants and at most five independent elements. In
this case, a matrix $M$ is always expressed such 
that $M=JM_{\rm r}J'$ where $J$ and $J'$ are the diagonal phase
matrices and $M_{\rm r}$ contains only real parameters. Thus the
matrices $MM^\dagger$ and $M^\dagger M$ are diagonalized by real
orthogonal matrices, up to overall phase rotations corresponding 
to $\Phi$ or $\Phi''$ in (\ref{U}). On the other hand, the phases
factors in $\Phi$ of the up and down sectors generally contribute to
the CKM matrix elements. Thus the CKM mixing matrix is found to be
written as
\begin{gather}
  V_{\rm CKM} \,=\, O_u^{\rm T} P O_d, \label{CKM} \\
  P \,=\, \Phi_u^*\Phi_d, \qquad
  O_i \,=\, O_1{}_i\,O_2{}_i\,O_3{}_i, \nonumber
\end{gather}
where the subscripts $i=u,d$ label the up- and down-type quarks,
respectively. As mentioned above, two complex phases in $P$ play an
important role for reproducing CP-violating quantities since their
changes generically affect the CKM matrix elements. The numerical
results of such phase factors will be discussed in later sections.

The experimentally observable quantities in the quark Yukawa sector
are 3 mixing angles with 1 complex phase in the CKM matrix and 6 mass
eigenvalues. The measured values of the CKM matrix elements
are~\cite{PDG}
\begin{equation}
  |V_{\rm CKM}|=\left|\begin{pmatrix}
    V_{ud} & V_{us} & V_{ub} \\
    V_{cd} & V_{cs} & V_{cb} \\
    V_{td} & V_{ts} & V_{tb}
  \end{pmatrix}\right|\,= \begin{pmatrix}
    0.9739-0.9751 & 0.221-0.227 & 0.0029-0.0045 \\
    0.221-0.227 & 0.9730-0.9744 & 0.039-0.044 \\
    0.0048-0.014 & 0.037-0.043 & 0.9990-0.9992
  \end{pmatrix}.
  \label{CKMexp}
\end{equation}
The various experimental observations of CP-violating phenomena yield
the CP violation in the standard model, which are translated
to~\cite{PDG}
\begin{equation}
  J_{\rm CP} \,=\, (2.88 \pm 0.33) \times 10^{-5},
  \label{J}
\end{equation}
where $J_{\rm CP}$ denotes the reparameterization-invariant measure of
CP violation~\cite{Jcp}. This value corresponds to the
Kobayashi-Maskawa phase in the standard parameterization 
as $\delta_{\rm KM}=60^\circ\pm14^\circ$. Moreover the recent results
of studying the decay of the $B$ mesons to charmoniums
indicate~\cite{CP-B}
\begin{equation}
  \sin 2\phi_1/\beta \,=\, 0.726\pm 0.037,
  \label{sin2b}
\end{equation}
where $\phi_1\equiv\beta$ is one of the angle of the unitary triangle
for the $B$-meson system, which is defined 
as $\phi_1=\beta\equiv\arg(V_{cd}^*V_{cb}/V_{td}^*V_{tb})$. This is
the angle that is most precisely known at present and is expected to
provide the most stringent constraint. The current-quark masses at 
the $Z$-boson mass scale are evaluated~\cite{FK} including various
effects such as the QCD strong coupling factors and we obtain
\begin{equation}
\begin{array}{lcl}
  m_u \,=\, 0.000975-0.00260, && m_d \,=\, 0.00260-0.00520, \\
  m_c \,=\, 0.598-0.702, && m_s \,=\, 0.0520-0.0845 \\
  m_t \,=\, 170-180, && m_b \,=\, 2.83-3.04,
\end{array}
\end{equation}
in GeV unit. The most recent results of the Tevatron CDF 
and D$\emptyset$ experiments indicate the top quark mass which is a
bit larger than that quoted above~\cite{CDF}. However, if taken into
account, the analysis of mass matrix forms presented in this paper is
not significantly changed, since the most influential ingredients are
practically the masses of lighter generations. In the following
analysis, we use these experimental data as input parameters and
explore possible forms of quark mass matrices.

\section{The Minimal Asymmetric Matrices}

We would like to systematically search for the mass matrices of up
and down quarks which are consistent with the current experimental
data. Our analysis is based on possible zero elements in the mass
matrices. Namely, the aim of this paper is to investigate how small
number of non-vanishing matrix elements can account for the existing
data. Here the number of zeros means independently-vanishing elements
in a mass matrix. In particular, for a symmetric 
matrix, ``1 zero'' implies that a diagonal element or a pair of
off-diagonal elements in symmetric positions takes a negligibly small
value.

Let us first consider the mass matrix of up-type quarks. In the
present work we assume that the up-quark mass matrix $M_u$ is
symmetric. This assumption is motivated by grand unified theory, where
the left- and right-handed up quarks in one generation often belong to
the same multiplet of unified gauge symmetry, like $SU(5)$ and larger. 
In this case one obtains the identical mixing matrix for left- and
right-handed up quarks; $V_{uL}=V_{uR}\equiv V_u$.

It is first noticed that three independent, non-vanishing matrix
elements are needed to reproduce the observed mass eigenvalues of the
three-generation quarks. Moreover a determinant of mass matrix must
be nonzero. The minimal forms of matrices which satisfy these
criterions are found to coexist with at most three zeros, and the
independent matrices are given by the following three types;
\begin{itemize}
\item Mu1
\begin{equation}
  M_u \,=\, \begin{pmatrix}
    a & & \\
    & b & \\
    & & c 
  \end{pmatrix}
\end{equation}
\item Mu2
\begin{equation}
  M_u \,=\, \begin{pmatrix}
    a & & \\
    & & b \\
    & b & c
  \end{pmatrix}
\end{equation}
\item Mu3
\begin{equation}
  M_u \,=\, \begin{pmatrix}
    & a & \\
    a & & b \\
    & b & c
  \end{pmatrix}
\end{equation}
\end{itemize}
The matrix elements $a$, $b$, $c$ are nonzero and the blanks denote
vanishing entries. All other forms of matrices consistent to the
criterions can be obtained by relabeling the generation indices. For
example, by exchanging the first and second generations (both for the
left- and right-handed fermions), the matrix Mu2 is converted to
the form discussed in~\cite{Giudice}
\begin{equation}
  M_u \,=\, \begin{pmatrix}
    & & b \\
    & a & \\
    b & & c
  \end{pmatrix}. \nonumber
\end{equation}
It should be noted that we do not assume any hierarchical orders
among the non-vanishing elements. Therefore the analysis of the above
three types of matrices (Mu1, Mu2, and Mu3) includes the whole
possibility of symmetric matrix with three zeros. The matrix $M_u$ is
diagonalized as
\begin{equation}
  M_u \,=\, V_u\begin{pmatrix}
  m_u & & \\
  & \!-m_c & \\
  & & m_t
  \end{pmatrix} V_u^\dagger.
\end{equation}
The negative sign in front of $m_c$ is just a convention introduced in
order that the mass eigenvalues and the parameter $c$ are real and
positive. With a suitable phase redefinition of the up-type quarks, we
take the non-vanishing matrix elements to be real parameters without
loss of generality. These values can be fixed by the three mass
eigenvalues from the following three equations:
\begin{eqnarray}
  \mathrm{tr}\,M_u &=& m_u - m_c + m_t,\\
  \mathrm{tr}\,M_u^2 &=& m_u^2 + m_c^2 + m_t^2,\\
  \det M_u &=& -m_u m_c m_t.
\end{eqnarray}
Thus a unitary matrix $V_u$ which diagonalizes a three-zero 
symmetric $M_u$ is described in terms of the up-quark mass eigenvalues.

For the down sector, the mass matrix is not necessarily symmetric. 
The minimal criterion for a realistic mass matrix is the same as for
symmetric matrices; three independent, non-vanishing matrix elements,
and a non-vanishing determinant. To satisfy these requirements, we
have at most six zero elements. A matrix with six zeros has only three
parameters which correspond to three mass eigenvalues. That only gives
no mixing angle or exchanging generation indices. However since any
type of up-quark mass matrices with three zeros cannot be diagonalized
with the observed CKM matrix, a matrix with asymmetric six zeros is
not suitable for the down sector. We thus find that the most
economical candidates for a realistic mass matrix of down-type quarks
have five zeros. They can generically describe three eigenvalues and
one mixing angle. It is found that there are 36 types of mass matrices
with five zeros and non-vanishing determinants. At this stage, since
we are not requiring any hierarchy among matrix elements as in the
case of up-quark mass matrices, these 36 ($=6\times 6$) patterns are
related to each other through the permutations of three rows and/or
three columns. Namely, one can obtain all the patterns by exchanging
the generation labels from a single matrix, e.g.,
\begin{equation}
  M_d \,=\, \begin{pmatrix}
   d & & \\
   & e & \\
   & g & f
  \end{pmatrix}.
\end{equation}
The matrix elements $d,\cdots,g$ are made real-valued by phase
redefinitions of quark fields. In our convention, a permutation of
columns corresponds to a rotation of generation indices of the
right-handed quarks, which rotations do not change the mass spectrum
and the CKM matrix elements. On the other hand, the exchanges of rows,
i.e.\ relabeling three left-handed down quarks, do affect on the
observable mixing angles. This is because we have already used the
label exchange degrees of freedom to reduce the number of matrix
patterns for the up-type quarks. Therefore all possible permutations
of rows must be taken into account in the down sector to explore the
whole combinations of up- and down-quark mass matrices. We thus
consider the following 6 types of mass matrices as the minimal
candidates with five zeros;
\begin{itemize}
\item Md1
\begin{equation}
  M_d \,=\, \begin{pmatrix}
    d & & \\
    & e & \\
    & g & f
  \end{pmatrix}
\end{equation}
\item Md2
\begin{equation}
  M_d \,=\, \begin{pmatrix}
    d & & \\
    & e & g \\
    & & f
  \end{pmatrix}
\end{equation}
\item Md3
\begin{equation}
  M_d \,=\, \begin{pmatrix}
    d & & \\
    g & e & \\
    & & f
  \end{pmatrix}
\end{equation}
\item Md4
\begin{equation}
  M_d \,=\, \begin{pmatrix}
    d & g & \\
    & e & \\
    & & f
  \end{pmatrix}
\end{equation}
\item Md5
\begin{equation}
  M_d \,=\, \begin{pmatrix}
    d & & \\
    & e & \\
    g & & f
  \end{pmatrix}
\end{equation}
\item Md6
\begin{equation}
  M_d \,=\, \begin{pmatrix}
    d & & g \\
    & e & \\
    & & f
  \end{pmatrix}
\end{equation}
\end{itemize}
As expected, these 6 patterns are transformed to each other by
changing the generation indices of the left-handed quarks, up to
permutations of the right-handed ones.

Given the possible forms of mass matrices, Mu1--Mu3 and Md1--Md6, we
analyze which combinations of mass matrices explain the observed
masses and mixing angles. It is easily found that the matrices Mu1 and
Mu2 cannot fit the data. This is because they have zero or one finite
mixing in the up sector and all the candidates of $M_d$ can induce
only one generation mixing, that necessarily results in the CKM matrix
with more than one vanishing entries. The only remaining possibility
is the matrix Mu3 for the up sector. The matrix which diagonalizes Mu3
is approximately written by the mass eigenvalues
\begin{equation}
  |O_u| \,\sim\, \begin{pmatrix}
    1 & \sqrt{\frac{m_u}{m_c}} & 
    \frac{m_c}{m_t}\!\sqrt{\frac{m_u}{m_t}} \\
    \sqrt{\frac{m_u}{m_c}} & 1 & \sqrt{\frac{m_c}{m_t}} \\
    \sqrt{\frac{m_u}{m_t}} & \sqrt{\frac{m_c}{m_t}} & 1  
  \end{pmatrix}.
\end{equation}
This shows that the 1-2 mixing angle from the up sector is roughly
given by $\sqrt{m_u/m_c}=0.037-0.066$. Therefore in order to generate
the observed Cabibbo angle, a 1-2 mixing angle from the down sector is
required to be of order $\mathcal{O}(10^{-1})$, which selects out Md3
or Md4 for an appropriate matrix for the down quarks. The matrices Md3
and Md4 are diagonalized by rotations of the first and second
generations and do not affect the third column 
of $V_{\rm CKM}$. Consequently the combinations (Mu3,$\,$Md3) 
and (Mu3,$\,$Md4) predict the CKM angles involving the third generation
\begin{eqnarray}
  &&|V_{ub}| \,\simeq\, \sqrt{\frac{m_u}{m_t}}\,=\, 0.00233-0.00391, \\
  &&|V_{td}| \,\simeq\, \bigg(|V_{us}|-\sqrt{\frac{m_u}{m_c}}\bigg)
  \sqrt{\frac{m_c}{m_t}}\,=\, 0.00894-0.0122, \\
  &&|V_{cb}| \,\simeq\, |V_{ts}| \,\simeq\, 
  \sqrt{\frac{m_c}{m_t}} \,=\, 0.0576-0.0643.
\end{eqnarray}
It is found that the first two predictions well agree with the
experimental values (\ref{CKMexp}) but the mixing between the second
and third generations is slightly larger than the observation.

We have found in this section that the simplest and realistic forms of
quark mass matrices can accommodate symmetric three zeros and
asymmetric five zeros in the up and down sectors, respectively. Their
explicit forms are highly constrained by the experimental values of
mass eigenvalues and mixing angles, and there exists only two
possibilities which are given by
\begin{equation}
  M_u \,=\, \begin{pmatrix}
    & a & \\
    a & & b \\
    & b & c
  \end{pmatrix}, \quad\quad 
  M_d \,=\, \begin{pmatrix}
    d & g & \\
    & e & \\
    & & f
  \end{pmatrix},
  \label{5zero1}
\end{equation}
and
\begin{equation}
  M_u \,=\, \begin{pmatrix}
    & a & \\
    a & & b \\
    & b & c
  \end{pmatrix}, \quad\quad 
  M_d \,=\, \begin{pmatrix}
    d & & \\
    g & e & \\
    & & f
  \end{pmatrix}.
  \label{5zero2}
\end{equation}
It is clear that a simultaneously exchange of the identical generation
indices for $u_L$, $u_R$, and $d_L$ completely preserves the physical
consequences. Moreover, as noted in the classification, any
permutation of the right-handed down quarks (i.e.\ of the columns 
of $M_d$) is also allowed phenomenologically. Noticing this fact, one
can see, for instance, that the matrix $M_d$ in (\ref{5zero2})
reconciles the Georgi-Jarlskog ansatz~\cite{GJ}, up to unphysical
field rotations, and could easily be extended to including the
charged-lepton mass matrix. A numerical evaluation for the combination
(\ref{5zero1}) presents us an example
\begin{eqnarray}
  M_u &=& \begin{pmatrix}
    0 & 0.000221\,(\lambda^{5.56}) & 0 \\
    0.000221\,(\lambda^{5.56}) & 0 & 0.0578\,(\lambda^{1.88}) \\
    0 & 0.0578\,(\lambda^{1.88}) & 0.997\,(\lambda^{0.00})
  \end{pmatrix}\!m_t, \\
  M_d &=& \begin{pmatrix}
    0.00156\,(\lambda^{4.27}) & 0.00518\,(\lambda^{3.48}) & 0 \\
    0 & -0.0285\,(\lambda^{2.35}) & 0 \\
    0 & 0 & 1.00\,(\lambda^{0.00})
  \end{pmatrix}\!m_b,
\end{eqnarray}
where the couplings are chosen so that we can fit as many observables
as possible to the experimental data. A non-trivial phase factor is
also required, e.g.\ $P=\textrm{diag.}(e^{-0.7\pi i},1,1)$, in order to
reproduce the observed value of CP violation~(\ref{J}). In this case,
however, we find that alternative indication of CP violation
(\ref{sin2b}) cannot be reproduced. The unitary triangle for 
the $B$-meson system is distorted due to a large value 
of $|V_{cb}|$, while the area of the triangle is correct. A similar
result is obtained for the combination (\ref{5zero2}) since physical
consequences are now determined modulo right-handed mixing of down
quarks. In the above example, we have alternatively written down in
the parentheses the exponents of a small 
parameter $\lambda$ ($=0.22$) for non-vanishing elements 
in $M_u$ and $M_d$. Such expressions in terms of an expansion
parameter would be suitable to gain an insight into the fermion masses
problem in a view of flavor symmetries. 

Though there is one unsatisfied point for the above two combinations
of mass matrices, i.e.\ a slightly large value of the 2-3 CKM mixing
angle ($|V_{cb}|\simeq\sqrt{m_c/m_t}$), one can easily find some
remedies. The discrepancy may be removed with radiative corrections,
for instance, the renormalization-group effects between the
electroweak and high-energy scales. A probable source of such
effects is a flexibility of the top-quark Yukawa coupling in
high-energy regime due to its fixed-point behavior in the infrared. To
see how the renormalization-group evolution makes the situation
better, let us first define the following ratio:
\begin{equation}
  \Delta \,\equiv\, \frac{\sqrt{m_c/m_t}}{|V_{cb}|}.  
\end{equation}
The problem with the mass matrices (\ref{5zero1}) and (\ref{5zero2})
is that their prediction $\Delta_{\rm pre}\simeq1$ is not consistent
to the experimental results $\Delta_{\rm exp}>1.31$ at the electroweak
scale. Within a good approximation that the third-generation Yukawa
couplings are dominant, the evolution of the mass ratio $m_c/m_t$ and
the mixing angle $V_{cb}$ is governed by the equations
\begin{eqnarray}
  \frac{d\ln\big(\frac{m_c}{m_t}\big)}{dt} &=& \frac{-1}{16\pi^2} 
  \big[\zeta({Y_u}_{33})^2 +\eta({Y_d}_{33})^2\big], \\
  \frac{d\ln|V_{cb}|}{dt} &=& \frac{-\eta}{16\pi^2}
  \big[({Y_u}_{33})^2 + ({Y_d}_{33})^2\big],
\end{eqnarray}
where $t=\ln\mu$ denotes the renormalization scale. The direct
contribution from gauge couplings is generally irrelevant to the
running of mass ratios and mixing angles. The 
coefficients $\zeta$ and $\eta$ are model-dependent constants, for
example, $\zeta=3/2$ and $\eta=-3/2$ in the standard 
model, and $\zeta=3$ and $\eta=1$ in supersymmetric standard 
models (though the ${Y_d}_{33}$ effect is negligible in the standard
model). Thus we obtain
\begin{equation}
  \frac{d\ln\Delta}{dt} \,=\, \frac{1}{32\pi^2}
  \big[(2\eta-\zeta)({Y_u}_{33})^2 +\eta({Y_d}_{33})^2\big].
  \label{rge}
\end{equation}
It is noted that the coefficient in front of $({Y_u}_{33})^2$ is
negative in usual Higgs doublet models. This negative sign suggests
that the ratio $\Delta$ in high-energy regime is reduced from the
value observed at the electroweak scale. Therefore the
renormalization-group evolution in fact ameliorates the problem with
the mass textures (\ref{5zero1}) and (\ref{5zero2}). It is also
noticed that, since the second term in the right-handed side 
of (\ref{rge}) is positive for supersymmetric cases, a smaller value
of ${Y_d}_{33}$ is preferred to cure the mismatch 
between $\Delta_{\rm pre}$ and $\Delta_{\rm exp}$. In the case that
only the top Yukawa coupling is dominant, we obtain by 
integrating (\ref{rge}) over the range between $\mu=1$ TeV and the
unification scale $\Lambda$,
\begin{equation}
  \frac{\Delta(\Lambda)}{\Delta(\mu)} \,=\,
  \bigg(\frac{{Y_u}_{33}(\Lambda)}{{Y_u}_{33}(\mu)}
  \bigg)^{\!\frac{-1}{12}}
  \!\bigg(\frac{g_1(\Lambda)}{g_1(\mu)}\bigg)^{\!\frac{-13}{1188}}
  \!\bigg(\frac{g_2(\Lambda)}{g_2(\mu)}\bigg)^{\!\frac{-1}{4}}
  \!\bigg(\frac{g_3(\Lambda)}{g_3(\mu)}\bigg)^{\!\frac{4}{27}}
  \;\simeq\; 0.91\,
  \bigg(\frac{{Y_u}_{33}(\Lambda)}{{Y_u}_{33}(\mu)}
  \bigg)^{\!\frac{-1}{12}}
\end{equation}
for the minimal supersymmetric standard model. In the second equation,
we have roughly assumed that the three gauge couplings of the standard
gauge groups $g_{1,2,3}$ are unified at $\Lambda$ in one-loop order
estimation. The situation is more improved for the standard model
because of a larger negative value of the 
coefficient $2\eta-\zeta$ and the negligible positive contribution
from the bottom Yukawa coupling, and one typically obtains
\begin{equation}
  \frac{\Delta(\Lambda)}{\Delta(\mu)} \,\sim\, 0.59\,
  \bigg(\frac{{Y_u}_{33}(\Lambda)}{{Y_u}_{33}(\mu)}
  \bigg)^{\!\frac{-1}{2}}
\end{equation}
at a high-energy scale $\Lambda\sim 10^{14}$ GeV\@. It is clearly
seen that $|V_{cb}|$ and $\sqrt{m_c/m_t}$ at high-energy regime become
closer than around the electroweak scale, and the texture
ansatz (\ref{5zero1}) and (\ref{5zero2}) would work better in
high-energy theory such as grand unified models.
   
On the other hand, a more direct resolution to the 
problem $\Delta_{\rm pre}\neq\Delta_{\rm exp}$ is to incorporate
additional non-vanishing elements into the Yukawa matrices. This is
the option we will explore in the next section.

\section{The Next-to-Minimal Asymmetric Matrices}

The previous analysis has shown that, at classical level, any
combinations of symmetric three-zeros $M_u$ and asymmetric five-zeros
$M_d$ are too simplified to be totally consistent with the observed
data. In this section we investigate a possibility to relax the
constraints on matrix forms and to introduce one more non-vanishing
matrix element.

The first case to consider is to work with symmetric two zeros in the
mass matrices of up quarks. An interesting observation is that, when
the 2-2 element in $M_u$ is turned on, the five-zeros 
matrices $M_d$ in (\ref{5zero1}) and (\ref{5zero2}) may completely
explain the data. This is because the mixing angle $V_{cb}$ could be
controlled by a free parameter in $M_u$, irrespectively of the charm
quark mass, which resolves the difficulty discussed in the previous
section. The zero structures of such mass matrices are ruled out at 
the $3\sigma$ level by the current experimental data if non-vanishing
elements are symmetric and hierarchical valued~\cite{KRS}. Such a
special case is obtained from our general form by exchanging the first
two indices of right-handed down quarks in (\ref{5zero2}) and
identifying the 1-2 and 2-1 matrix elements. Up to relabeling
generation indices, there are 4 types of mass textures with symmetric
two zeros. Exploring all possible patterns for $M_u$ (with symmetric
two zeros) and $M_d$ (with asymmetric five zeros), we find that the
following mass matrices are successful to explain the experimental data:
\begin{align}
  \textrm{\fbox{$M_u$}} && \textrm{\fbox{$M_d$}} \nonumber \\
  & \begin{pmatrix}
    & a & \\
    a & d & b \\
    & b & c
  \end{pmatrix} &&
  \begin{pmatrix}
    e & h & \\
    & f & \\
    & & g
  \end{pmatrix} \nonumber\\
  & \begin{pmatrix}
    & & a \\
    & b & d \\
    a & d & c
  \end{pmatrix} &&
  \begin{pmatrix}
    e & & \\
    h & f & \\
    & & g
  \end{pmatrix} \nonumber
\end{align}
All $4=2\times2$ combinations of $M_u$ and $M_d$ well describe the
present experimental data. It should be noted that there are
additional solutions with relabeling the generation indices, while
physical implications are unchanged. First, any permutation 
of $d_R$ (i.e.\ of the columns of $M_d$) is phenomenologically
allowed. In addition, simultaneously exchanges of the identical
generation indices for $u_L$, $u_R$, and $d_L$ completely preserves
the physical consequences and also become the solutions. The above
sets of mass matrices are consistent with all the properties of
quarks, including the recent measurements of CP violation in 
the $B$-meson system (\ref{sin2b}).

The other case we will pursue in the following is to extend the
down-type mass matrix $M_d$ to contain (asymmetric) four zeros. It is
found that there are 81 types of matrices with nonzero determinants,
which are related through the permutations of rows and/or columns. All
possible forms can be generated from the following four
representatives of mass matrices:
\begin{equation}
\begin{pmatrix}
  d & & h \\
  & e & g \\
  & & f
\end{pmatrix} \qquad
\begin{pmatrix}
  d & & \\
  & e & \\
  h & g & f
\end{pmatrix} \qquad
\begin{pmatrix}
  & e & \\
  d & & h \\
  & g & f
\end{pmatrix} \qquad
\begin{pmatrix}
  d & & h \\
  & e & \\
  g & & f
\end{pmatrix}
\label{repr}
\end{equation}
It is first noticed that the last matrix generates only one generation
mixing angle and does not cure the problem in the previous section,
where one of the CKM mixing angles is not entirely consistent to the
experimental data. Therefore we can safely drop this matrix (and the
other 8 ones generated by changing the labels) in the analysis
below. For the former two mass matrices in (\ref{repr}), a permutation
of the first two columns produces the same modification as that
obtained by permuting the first two rows, since any hierarchical order
among the matrix elements is not supposed at this stage. This fact
reduces by half the number of independent forms of matrices generated
by exchanging the generation labels from the first two representatives
in (\ref{repr}). The total number of independent mass matrices we will
consider is $72$ ($=18+18+36$).

As in the analysis of Section~3, if one does not count up the
rotations of the right-handed down-type quarks (the columns of
matrices), only the following 12 types of matrices should be taken
into account:
\begin{eqnarray}
\begin{pmatrix}
  d & & h \\
  & e & g \\
  & & f
\end{pmatrix}\,
\begin{pmatrix}
  d & h & \\
  & e & \\
  & g & f
\end{pmatrix}\,
\begin{pmatrix}
  d & & \\
  h & e & \\
  g & & f
\end{pmatrix}\,
\begin{pmatrix}
  d & & \\
  & e & \\
  h & g & f
\end{pmatrix}\,
\begin{pmatrix}
  d & & \\
  g & e & h \\
  & & f
\end{pmatrix}\,
\begin{pmatrix}
  d & h & g \\
  & e & \\
  & & f
\end{pmatrix}\nonumber\\
\begin{pmatrix}
  & e & \\
  d & & h \\
  & g & f
\end{pmatrix}\,
\begin{pmatrix}
  d & & h \\
  & e & \\
  & g & f
\end{pmatrix}\,
\begin{pmatrix}
  d & & \\
  h & e & \\
  & g & f
\end{pmatrix}\,
\begin{pmatrix}
  d & & h \\
  g & e & \\
  & & f
\end{pmatrix}\,
\begin{pmatrix}
  d & h & \\
  & e & g \\
  & & f
\end{pmatrix}\,
\begin{pmatrix}
  d & h & \\
  & e & \\
  g & & f
\end{pmatrix}
\label{12}
\end{eqnarray}
Since we now drop the degrees of freedom of the right-handed down
rotations, physical consequences can be read from the symmetric 
matrix $M_dM_d^\dagger$, which has 6 independent elements. For the
above 12 types, a hermitian matrix $M_dM_d^\dagger$ contains 5 free
parameters and necessarily leads to one vanishing element or one
relation among non-vanishing elements. Thus an asymmetric four-zeros
matrix gives similar results to that of a symmetric one-zero matrix,
as far as mass eigenvalues and left-handed down mixing are
concerned. It is noticed that there is a difference 
between $M_dM_d^\dagger$ and symmetric mass matrices discussed
in~\cite{RRR} that none of the diagonal elements 
of $M_dM_d^\dagger$ can be zero for matrices with non-vanishing
determinants. It was found in the analysis of Ref.~\cite{RRR} that
symmetric two-zeros mass matrices for the down sector are consistent
with the experimental data. Given these facts, the above asymmetric
four-zeros mass matrices are expected to account for the proper mass
eigenvalues and mixing, because they correspond to symmetric one-zero
matrices which have one more free parameter. The number of
combinations which can explain the data is hence rather large, and the
exploration along this line unfortunately seems not to provide a new
perspective for the origin of fermion masses and mixing angles.

Let us proceed the discussion taking account of the roles played by
the mixing of right-handed down quarks. In the standard model, the
mixing of right-handed ($SU(2)$-singlet) fermions is irrelevant to the
CKM mixing and unphysical (unobservable) degrees of freedom. This is
not necessarily true in various extensions of the standard model. For
example, in supersymmetric extensions of the standard model, the
right-handed mixing of fermions which diagonalizes a Yukawa matrix is
transfered to that of corresponding scalars via supersymmetry-breaking
scalar masses. Thus the masses of scalar superpartners generally have
generation dependences and cause observable effects, such as
flavor-changing decays of heavy fermions. A more interesting situation
arises in the frameworks of grand unification. In this case, quarks
and leptons are unified into some multiplets of unified gauge group
and the mass matrices of quarks are often closely related to those of
leptons. This fact may give rise to an apparent difficulty in
simultaneously realizing the small CKM mixing and the observed large
lepton mixing, which is described by the 
Maki-Nakagawa-Sakata (MNS) matrix $V_{\rm MNS}$~\cite{MNS}. The
parallelism between quarks and leptons, which is a sign of grand 
unification, does not seem to work in the Yukawa sector. There is
however an interesting observation that the mixing angles of
left-handed charged leptons are correlated to those of right-handed
down quarks and therefore the CKM mixing does not necessarily connect
with the MNS mixing. This idea is easily achieved in $SU(5)$ grand
unification and larger unified theories~\cite{lopsided}, where a key
ingredient is that an anti 5-plet of $SU(5)$ contains one-generation
right-handed down quark and left-handed lepton doublet, and also there
are some multiplicities of anti 5-plets, which are naturally
incorporated in $SO(10)$ or $E_6$ unified models. This mechanism
automatically makes $M_d$ asymmetric since it changes the property of
anti 5-plets only, while keeping that of 10-plets.

In this way the mixing of right-handed fermions is not necessarily
unobservable quantities. In the following analysis, motivated by grand
unification view mentioned above, we investigate possible connections
between $V_d{}_R$ and the leptonic mixing matrix which diagonalizes
the mass matrix of three-generation charged leptons. In particular, we
examine which combinations of mass matrices for the up and down
sectors suggest large leptonic mixing, recently observed in various
neutrino experiments. Since the flavor rotation of right-handed down
quarks is now supposed to change physical consequences, it is not
used to reduce the number of candidates for four-zero matrices as done
in the above. We will therefore exhaust all of the most generic 72
candidates for realistic down-quark mass matrices with four vanishing
entries.

Assuming that the neutrino oscillations account for the solar and
atmospheric neutrino data, the recent experimental results indicate
rather large angles for the 1-2 and 2-3 generation mixing in the MNS
matrix, but a small one for the 1-3 mixing 
angle $(V_{\rm MNS})_{13}<0.14-0.22$~\cite{CHOOZ}. As for the two
large mixing angles, the best fit value for the atmospheric neutrino
data is the maximal mixing ($\theta\simeq\pi/4$), and on the other
hand, the solar neutrino deficit needs a large but non-maximal value
of the 1-2 mixing angle~\cite{analyses}. In the following analysis, we
first consider, just as a first approximation, the maximal angles both
for the 1-2 and 2-3 mixing, and then examine possible deviations from
these maximal angles. The lepton mixing matrix is defined as
\begin{equation}
  V_{\rm MNS} \,=\, V_{eL}^\dagger V_\nu,  
\end{equation}
where $V_{eL}$ rotates the three-generation charged leptons such that
the mass matrix of charged leptons is 
diagonalized, and $V_\nu$ denotes some mixing matrix in the neutrino
sector. Its form crucially depends on the neutrino property and we 
leave it, together with detailed analysis of neutrino mass texture
zeros, to another future task~\cite{next}. It should be noted that
this does not mean that we take $V_\nu=1$ in the following
analysis. In fact, our result will show that considerable contribution
to lepton mixing is needed to come from neutrino mass matrices, which
could be realized in a huge variety of neutrino models.

As we mentioned, if embedding the theory into grand unification
scheme, the mixing of charged leptons $V_{eL}$ may be related to that
of right-handed down quarks as $V_{eL}\simeq V_{dR}$, up to
corrections due to the breaking of unified gauge symmetry. To
precisely reproduce the mass eigenvalues of charged leptons, it is in
fact needed to take in some breaking effects which split the
properties of quarks and leptons. Typical examples of such splitting
are provided by the Georgi-Jarlskog  factor~\cite{GJ} and
higher-dimensional operators involving Higgs fields that break
quark-lepton symmetry. We assume, just for simplicity, that such
breaking effects are small not to significantly change the analysis
below. We are thus interested in the following typical forms of mixing
matrices for the down sector, which are associated with large
generation mixing of left-handed charged leptons:\footnote{The 
mixture (\ref{solar}) would be excluded by the neutrino oscillation 
experiments as it would generate too a large value of the 1-3 lepton mixing
angle, if the atmospheric neutrino angle comes from the neutrino
sector. While included into the analysis, as we will show, the mixing
pattern (\ref{solar}) is already disfavored by the quark data alone.}
\begin{eqnarray}
V_{dR} &=& \begin{pmatrix}
  1 & 0 & 0 \\
  0 & 1/\sqrt{2} & -1/\sqrt{2} \\
  0 & 1/\sqrt{2} & 1/\sqrt{2}
\end{pmatrix},
\label{atm}\\
V_{dR} &=& \begin{pmatrix}
  1/\sqrt{2} & -1/\sqrt{2} & 0 \\
  1/\sqrt{2} & 1/\sqrt{2} & 0 \\
  0 & 0 & 1
\end{pmatrix},
\label{solar}\\
V_{dR} &=& \begin{pmatrix}
  1/\sqrt{2} & -1/2 & 1/2 \\
  1/\sqrt{2} & 1/2 & -1/2 \\
  0 & 1/\sqrt{2} & 1/\sqrt{2}
\end{pmatrix}.
\label{bilarge}
\end{eqnarray}

A general procedure for examining viable forms of $V_{dR}$ is as
follows. At first, evaluate the matrix
\begin{equation}
  M_d^\dagger M_d \,=\, V_{dR} \begin{pmatrix}
    m_d^2 & & \\
    & m_s^2 & \\
    & & m_b^2
  \end{pmatrix} V_{dR}^\dagger,
\label{VdR}
\end{equation}
where the matrix $V_{dR}$ is parameterized as given in Section~2. We
consider in this section the matrices $M_d$ with asymmetric four
zeros. Namely, they contain five free parameters, and a matrix of the
form $M^\dagger M$ has six independent elements. The 
equation (\ref{VdR}) therefore imposes one constraint which can be
used to eliminate a mixing angle of the right-handed down quarks. As
noted above, one or two additional constraints are obtained to reduce
the number of independent (mixing) parameters, when one explores the
solutions of $V_{dR}$ with large mixing. For example, we can 
fix $(V_{dR})_{32}=1/\sqrt{2}$ in the case of (\ref{atm}). Once the 
matrix elements in $M_d$ are solved with respect to remaining
independent parameters, the mixing matrix for left-handed down quarks
is expressed as
\begin{equation}
  V_{dL} \,=\, M_d V_{dR} \begin{pmatrix}
  m_d^{-1} & & \\
  & \!\!m_s^{-1} & \\
  & & \!\!m_b^{-1}
\end{pmatrix}.
\end{equation}
Such left-handed down mixing is used to examine which forms of mass
matrices produce the observed values of the CKM matrix elements.

Numerically exhausting all the possible forms of mass matrices, we
find that the down-quark mass matrices with asymmetric four zeros are
unfavorable to a sizable value of mixing angle between the first and
second generations of right-handed down quarks. This fact is easily
understood in the case that there is large mixing only between the
first two generations. Such a mixing matrix is given by
\begin{equation}
  V_{dR} \,=\, \begin{pmatrix}
  \cos\theta & -\sin\theta & 0 \\
  \sin\theta & \cos\theta & 0 \\
  0 & 0 & 1
\end{pmatrix}
\label{VdR2}
\end{equation}
at the leading order of other small mixing angles. The case 
where $\theta\sim\pi/4$ corresponds to Eq.~(\ref{solar}) and could
provide a solution to the solar neutrino problem in grand unification
schemes. To explain the quark mixing angles for the up-quark matrices
Mu1, Mu2 and Mu3, the left-handed mixing in the down sector needs to
satisfy
\begin{equation}
  V_{dL} \,\simeq\, \begin{pmatrix}
    1 & \mathcal{O}(\epsilon) & \mathcal{O}(\epsilon^3) \\
    \mathcal{O}(\epsilon) & 1 & \mathcal{O}(\epsilon^2) \\
    \mathcal{O}(\epsilon^3) & \mathcal{O}(\epsilon^2) & 1
\end{pmatrix},
\label{VdL}
\end{equation}
where $\epsilon$ is a small parameter of order of $10^{-1}$. It is found
from the analysis in the previous section that the marginal
requirements are sizable contributions from the down sector 
to $V_{us}$ and $V_{cb}$ (and not necessarily to $V_{ub}$). This is
translated to lower bounds on the left-handed mixing of down quarks,
for example, $|(V_{dL})_{12}|>0.16$ and $|(V_{dL})_{23}|>0.012$. When
there is a solution for the above-described procedure, the
corresponding down-quark mass matrix is given by
\begin{equation}
  M_d \,\simeq\, 
  \begin{pmatrix}
    1 & \mathcal{O}(\epsilon) & \mathcal{O}(\epsilon^3) \\
    \mathcal{O}(\epsilon) & 1 & \mathcal{O}(\epsilon^2) \\
    \mathcal{O}(\epsilon^3) & \mathcal{O}(\epsilon^2) & 1
  \end{pmatrix}
  \begin{pmatrix}
    m_d & & \\
    & m_s & \\
    & & m_b
  \end{pmatrix}
  \begin{pmatrix}
    \cos\theta & \sin\theta & 0 \\
    -\sin\theta & \cos\theta & 0 \\
    0 & 0 & 1
  \end{pmatrix}.
  \label{Md1}
\end{equation}
It is clearly seen that four zeros in $M_d$ cannot be realized since
the matrix elements in the second and third rows are always
non-vanishing for any precise values of $V_{dL}$ (\ref{VdL})
satisfying the lower bounds mentioned in the above. The situation
might be improved by turning on fluctuations around the exact form 
of $V_{dR}$ (\ref{VdR2}). In this case, one is in fact able to take
either $(M_d)_{31}$ or $(M_d)_{32}$ as zero, if 
additional $\mathcal{O}(\epsilon^4)$ mixing in $V_{dR}$ is
introduced. However not all of the elements in the first row become
zero, in particular, either $(M_d)_{11}$ or $(M_d)_{12}$ can be set to
zero. We thus find that the down-quark mass matrices with asymmetric
four zeros generically conflict with large 1-2 mixing in the
right-handed down sector.

It turns out that solutions with two large mixing 
like (\ref{bilarge}) are also absent. In the limit of bi-maximal
mixing of $d_R$, the down-quark mass matrix becomes
\begin{equation}
  M_d \,\simeq\, 
  \begin{pmatrix}
    1 & \mathcal{O}(\epsilon) & \mathcal{O}(\epsilon^3) \\
    \mathcal{O}(\epsilon) & 1 & \mathcal{O}(\epsilon^2) \\
    \mathcal{O}(\epsilon^3) & \mathcal{O}(\epsilon^2) & 1
  \end{pmatrix}
  \begin{pmatrix}
    m_d & & \\
    & m_s & \\
    & & m_b
  \end{pmatrix}
  \begin{pmatrix}
    \frac{1}{\sqrt{2}} & \frac{1}{\sqrt{2}} & 0 \\
    \frac{-1}{2} & \frac{1}{2} & \frac{1}{\sqrt{2}} \\
    \frac{1}{2} & \frac{-1}{2} & \frac{1}{\sqrt{2}}
  \end{pmatrix}.
  \label{Md}
\end{equation}
The situation is rather different from the case of (\ref{Md1}), for
instance, it is now possible to have vanishing matrix elements in the
second row in (\ref{Md}) due to the presence of the third-generation
large mixing. Note that the third-row elements are necessarily
non-vanishing, even if one introduces sizable deviations from the
maximal or zero mixing angles 
in $V_{dR}$, i.e.\ $\theta_{1,3}^{\,d_R}\neq\pi/4$ and/or
$\theta_2^{\,d_R}\neq0$. Accordingly it is enough to consider physical 
consequences of the matrices $M_d$ with four zeros placed in the first
and second rows (and the other components are nonzero)~\cite{AFM}. If
one adopts the up-quark mass matrices Mu1--Mu3, the mixing angles from
the down sector have some lower bounds, in 
particular, $|(V_{dL})_{12}|>0.16$ is needed. It is numerically 
evaluated that the condition $|(V_{dL})_{12}|>0.16$ constrains the
other mixing angles as $|(V_{dR})_{21}|<0.40$ ($0.44$) 
for $|(V_{dR})_{32}|=0.7$ ($0.61$). This value is translated, in the
limit of negligible 1-3 mixing, to the upper bound of the 1-2 mixing 
angle $\theta_3<26.1^\circ$, which is excluded at more 
than $3\sigma$ level by the recent results of neutrino experiments, if
generation mixing in the neutrino sector is found to be small. We
thus find that any four-zero mass matrix in the down sector is not
compatible with bi-large generation mixing of right-handed down
quarks.

Finally let us consider the case that the right-handed down mixing
between the second and third generations is large and the others are
suppressed [Eq.~(\ref{atm})]. This implies, if adopting the grand
unification, the amount of mixture of the atmospheric
neutrinos. Following the general procedure described before, we have
exhausted the possible patterns and found that, at classical level,
the following mass matrices satisfy the criterion for (charged) lepton
mixing, while the quark masses and the CKM matrix elements (including
the KM phase) are properly reproduced:
\begin{align}
  \textrm{\fbox{$M_u$}} && \textrm{\fbox{$M_d$}} \nonumber \\
  & \begin{pmatrix}
    a & & \\
    & & b \\
    & b & c
  \end{pmatrix} &&
  \begin{pmatrix}
    d & e & \\
    & & h \\
    & g & f
  \end{pmatrix} \nonumber\\
  & \begin{pmatrix}
    & a & \\
    a & & b \\
    & b & c
  \end{pmatrix} &&
  \begin{pmatrix}
    & e & \\
    d & & h \\
    & g & f
  \end{pmatrix} \nonumber\\
  &&& \begin{pmatrix}
    & e & \\
    & & h \\
    d & g & f
  \end{pmatrix} \nonumber
\end{align}
All $6=2\times3$ combinations of $M_u$ and $M_d$ are consistent to the
present experimental data. Note that one type of mass texture 
for $M_d$ [the second one in (\ref{12})] almost describes the data we
have listed in Section~2. It is however found that the whole parameter
space of that texture is excluded by the measured value of CP
violation in the $K$-meson system (the $\epsilon_K$ constraint). All
the other combinations of mass textures are not compatible to the 
experimentally allowed parameter region for the atmospheric neutrino
problem at more than $6\sigma$ level, unless there is sizable
contribution to mixing angles from the neutrino sector. The examples
of numerical fits for these matrix elements are shown in the
appendix. It should be noted that there are additional solutions with
relabeling the generation indices, while physical implications are
unchanged. The combinations obtained by exchanging the second and
third columns of $M_d$ are viable. This is simply because we now
consider the situation that the second and third generations of $d_R$
are largely mixed. In addition, simultaneously exchanges of the
identical generation indices for $u_L$ (and $u_R$ for symmetric 
textures) and $d_L$ completely preserves the physical consequences and
also become the solutions. No other exchanging symmetry does not
exist. The above sets of mass matrices are consistent with all the
properties of quarks, including the recent measurements of CP
violation in the $B$-meson system, as well as large lepton mixing for
the atmospheric neutrino problem. More conservatively, they provide
sizable contribution to leptonic 2-3 mixing, while satisfying the
experimental results of quark masses and the CKM matrix elements. It
is interesting to see that, in the down sector, the numerical
exploration of parameter space shows that a 
correlation $f\simeq g$ should hold for all the above solutions. The 
above list of textures contains the Fritzsch ansatz~\cite{Fritzsch},
but $M_u$ is symmetric while $M_d$ is not~\cite{BS}, leading to large
generation mixing in $V_{dR}$. As mentioned in the beginning of this
section, such an asymmetric form of mass matrix often plays a key role
for neutrino physics in grand unified theory~\cite{lopsided}. We have
shown that the above three forms of $M_d$ are the minimal extensions
of the ansatz $f\sim g$ to including the first generation. The
obtained matrices are successful to explain the observed quark masses
and mixing angles and have the maximal number of vanishing elements.

\section{Summary and Discussions}

The study of the origin of fermion masses and mixing angles is one of
the most important unresolved issues in particle physics. As a
plausible approach to this issue, possible zero elements in mass
matrices have been extensively examined and the obtained results have
suggested useful guides for realistic model construction. In this
paper we have systematically investigated what types of quark mass
matrices with non-symmetrical forms can be consistent with the
experimentally obtained CKM matrix and mass eigenvalues. Our first
principle is that a mass matrix has as simple form as possible,
namely, to search for the minimal number of free parameters in the
mass matrices. This leads us to consider some of mass matrix elements
to be vanishing. The existence and structures of zero matrix elements
are expected to be deeply connected with underlying physics, such as
flavor symmetries, in more fundamental theory of quarks and
leptons. We have first examined experimentally viable mass matrices in
the case where the up-quark sector has symmetric three zeros and the
down-quark sector asymmetric five zeros. This is the simplest
possibility apparently not to conflict with the experimental data, and
can almost explain the observed quark masses and mixing angles. The
situation is rather different from the case where the down-quark mass
matrix contains at most four vanishing elements. We then find that
there exist various forms of mass matrices consistent with the
existing experimental data, and it seems difficult to find some clues
to understand the generation structure. Additional information comes
from the recent observation of neutrino generation mixing. If working
with the grand unification hypothesis, the mixing of $SU(2)$-doublet
leptons is correlated to that of $SU(2)$-singlet down-type quarks. To
investigate the implications of large mixing angles in the lepton
sector, we have searched viable solutions which induce large
right-handed mixing in the down sector, and found that there only
exist six patterns of mass matrices with a large mixture between the
second and third generation of right-handed down quarks. Furthermore
it turns out in our framework that the large angle solution for the
solar neutrino problem cannot be realized from the charged lepton
sector with asymmetric four-zeros $M_d$ ($M_e$).

The observed large amount of mixing angle of solar neutrinos then
should come from the neutrino sector. If the minimality principle is
applied to neutrino mass matrices, the simplest matrix 
forms (i.e.\ with the maximal number of zero matrix elements) could be
found out. However the neutrinos have rich phenomenology and their
property has not been fixed experimentally. In particular, there still
exists wide possibilities for the neutrino mass spectrum, which fact
generally makes the thorough analysis of neutrino mass textures
laborious. We here briefly discuss several results for possible forms
of neutrino mass matrices which lead to the large lepton mixing
between the first and second generations and have the maximal number
of allowed zero matrix elements. First, consider the effective
neutrino mass operator $\kappa_{ij}\bar L_iL_jH^*H$ where $L_i$ denote
the three-family lepton doublets. The coefficient 
matrix $\kappa_{ij}$ is symmetric in the generation space. This
higher-dimensional operator induces the Majorana neutrino mass 
matrix $M_L=\kappa\langle H^*H\rangle$ after the electroweak gauge
symmetry breaking. We find two types of the simplest forms 
of $M_L$ which have symmetric four zeros and are given by
\begin{equation}
  M_L \,=\, \begin{pmatrix}
    & l & \\
    l & n & \\
    & & ~
  \end{pmatrix} \quad{\rm or}\quad
  \begin{pmatrix}
    n & l & \\
    l & & \\
    & & ~
  \end{pmatrix}.
  \label{ML}
\end{equation}
It is interesting to note that this form of neutrino mass matrix
predicts the spectrum with the inverted mass hierarchy and an exactly
massless neutrino for the third generation. Moreover, taken into 
account the observed neutrino mass differences, such $M_L$ leads to
almost maximal mixing angle between the first two generations. As a
result, it can be matched with only four of six combinations of quark
mass matrices found in the previous section. If the minimality
analysis is extended to the next-to-minimal level, i.e.\ $M_L$ with
symmetric three zeros, we find thirteen patterns are allowed, each of
which predicts characteristic mass spectrum of light Majorana
neutrinos. Another well-known scheme for light neutrinos is to
consider ($3\times3$) Dirac neutrino mass matrix $M_\nu$ and
right-handed Majorana one $M_R$. Also in this case, the analysis is
quite different from the quark sector, mainly because one neutrino can
be massless. We find from the exhaustive exploration that the maximal
number of vanishing matrix elements is ten which consists of
asymmetric seven (six) zeros in $M_\nu$ and symmetric 
three (four) zeros in $M_R$. There are four patterns of the
seven-zeros $M_\nu$ cases, which contain as an example
\begin{equation}
  M_\nu \,=\, \begin{pmatrix}
    s & & \\
    & t & \\
    & & ~
  \end{pmatrix}, \quad\quad 
  M_R \,=\, \begin{pmatrix}
    & u & \\
    u & & v \\
    & v & w
  \end{pmatrix},
\end{equation}
and six patterns for the six-zeros $M_\nu$ cases, for example,
\begin{equation}
  M_\nu \,=\, \begin{pmatrix}
    & s & \\
    t & u & \\
    & & ~
  \end{pmatrix}, \quad\quad 
  M_R \,=\, \begin{pmatrix}
    & v & \\
    v & & \\
    & & w
  \end{pmatrix}.
\end{equation}
All the ten patterns of $M_\nu$ and $M_R$ generate light neutrino mass
matrix $M_L$ in the form of (\ref{ML}) after the seesaw
operation. Therefore the resultant mass spectrum and possible partners
for quark mass matrices are the same as the cases (\ref{ML}). Some
different phenomenology may appear through lepton flavor-violating
processes induced by lepton Yukawa couplings~\cite{FV}. The detailed
analysis of minimal lepton mass matrices and their phenomenological
implications will be presented in a separate paper~\cite{next}.

In the analysis of this paper, except for the discussion at the end of
Section 3, we have not taken into account of the dependence of matrix
forms on the renormalization scale, and considered generic features 
of 3$\times$3 quark Yukawa couplings including asymmetrical
matrices. For more precise treatment, the renormalization-group
evolution of Yukawa couplings are required to be evaluated, because
zeros of matrix elements should be implemented at some high-energy
scale such as a grand unification scale. The observable quantities at
the electroweak scale deviate to some extent from the values estimated
in high-energy regime. However one of the most important points is
that the fermion mass ratios of the first to second generations is
almost insensitive to radiative corrections due to the fact that the
dominant contribution to flavor-changing evolution comes from the
Yukawa couplings of the third generation. In our analysis, the
selection of viable forms of mass matrices has mainly depended on
whether the down-quark matrices satisfy the experimental value of the
1-2 CKM mixing in conspiracy with the up sector. It is therefore
expected that the renormalization-group analysis does not destabilize
the results of our analysis of possible zero elements, while there
certainly exist some scale dependences of non-vanishing matrix
elements in the presence of significant contributions from the gauge
and top-quark Yukawa couplings. This latter fact is supposed to only
change `initial' values of non-vanishing Yukawa couplings at a
high-energy scale. The results presented in this paper are also useful
for explaining the flavor structures of quarks and leptons in grand
unification schemes.

Finally, we would like to comment on some phenomenology related to the 
solutions obtained in Section~4. These solutions predict similar sizes
of off-diagonal elements to the 3-3 elements and radiative corrections
from Yukawa couplings are important for flavor physics. For 
example, if the theory is supersymmetrized, flavor violation in the
Yukawa sectors is translated to off-diagonal components of
supersymmetry-breaking scalar masses through the radiative
corrections. That could induce sizable rates of flavor-changing
neutral currents for quarks and charged leptons~\cite{FV} in
supergravity models. Further searches of flavor-violating processes
will provide us a new perspective of flavor structures in high-energy
regime.

In the viewpoint of distinguishing possible solutions, it is important 
to examine observable signals of underlying theory. In addition to 
signals of underlying symmetries or dynamics, the improved
measurements of low-energy observable quantities allow us to
discriminate discrete ambiguities of possible matrices. As for the
solutions 1--6 presented in the appendix, it can be seen from the
numerical analysis that the solutions 3 and 4 have sizable
contributions to $(V_{dR})_{31}$ components. This means that they 
predict $(V_{\rm MNS})_{13}\sim\mathcal{O}(10^{-1})$ if there appears 
no fine tuning of parameters in $V_{dR}$ and $V_\nu$. Since the
planned improvements in the sensitivity to $(V_{\rm MNS})_{13}$ are
expected to reach 0.05~\cite{neufact}, these solutions would be
supported or disfavored when a precise value 
of $(V_{\rm MNS})_{13}$ is measured. For other generation mixing, the
solutions 3--6 are found to have relatively larger values 
of $(V_{dR})_{21}$ (of the order of the Cabibbo angle) than the other
solutions. This fact also distinguishes possible textures, for
example, if the theory is extended to incorporate 
supersymmetry (breaking) or grand unification. Together with these
issues stated above and others, it is hoped to find what underlying
theory governs the masses and mixing angles of quarks and leptons.

\bigskip
\subsection*{Acknowledgments}

This work is supported by the grant-in-aid for scientific research on
priority areas (No.~441): "Progress in elementary particle physics of
the 21st century through discoveries of Higgs boson and supersymmetry"
(No.~16081209).

\newpage
\appendix
\section{The order estimation and numerical evaluations of the quark
mass matrices}

In this appendix, we would like to present the order estimation of
quark mass matrix elements and typical examples of numerical fitting
for the solutions obtained in Section~4, where the up-quark matrices
have symmetric three zeros and the down-quark ones asymmetric four
zeros. Since the observed values of masses and mixing angles are
hierarchical, one could parameterize matrix elements by integer
exponents of a small parameter $\lambda$ ($=0.22$) times $O(1)$
coefficients, which originate from the ambiguities of Yukawa coupling
constants. Such expressions with integer exponents might be useful for
getting ideas of constructing fermion mass matrix models with flavor
symmetries. We have found in Section~4 that there 
are $6=2\times3$ combinations of up and down quark mass textures well
describe the current experimental data. The order estimation of these
mass matrix elements are presented in Table~\ref{oe}, where we have
not explicitly written down $O(1)$ coefficients mentioned above.
\begin{table}[htbp]
\renewcommand{\arraystretch}{1.05}
\centering
\begin{tabular}{c|cc} \hline\hline
& $M_u\,/m_t$ & $M_d\,/m_b$ \\ \hline
1 & ~~$\begin{pmatrix}
  \lambda^8 & & \\
  & & \lambda^2 \\
  & \lambda^2 & \lambda^0
\end{pmatrix}$~~ &
~~$\begin{pmatrix}
  \lambda^4 & \lambda^3 & \\
  & & \lambda^2 \\
  & \lambda^0 & \lambda^0
\end{pmatrix}$~~ \\ \hline
2 & $\begin{pmatrix}
  & \lambda^6 & \\
  \lambda^6 & & \lambda^2 \\
  & \lambda^2 & \lambda^0
\end{pmatrix}$ &
$\begin{pmatrix}
  \lambda^4 & \lambda^3 & \\
  & & \lambda^2 \\
  & \lambda^0 & \lambda^0
\end{pmatrix}$ \\ \hline
3 & $\begin{pmatrix}
  \lambda^8 & & \\
  & & \lambda^2 \\
  & \lambda^2 & \lambda^0
\end{pmatrix}$ &
$\begin{pmatrix}
  & \lambda^3 & \\
  \lambda^3 & & \lambda^2 \\
  & \lambda^0 & \lambda^0
\end{pmatrix}$ \\ \hline
4 & $\begin{pmatrix}
  & \lambda^6 & \\
  \lambda^6 & & \lambda^2 \\
  & \lambda^2 & \lambda^0
\end{pmatrix}$ &
$\begin{pmatrix}
  & \lambda^3 & \\
  \lambda^3 & & \lambda^2 \\
  & \lambda^0 & \lambda^0
\end{pmatrix}$ \\ \hline
5 & $\begin{pmatrix}
  \lambda^8 & & \\
  & & \lambda^2 \\
  & \lambda^2 & \lambda^0
\end{pmatrix}$ &
$\begin{pmatrix}
  & \lambda^3 & \\
  & & \lambda^2 \\
  \lambda^1 & \lambda^0 & \lambda^0
\end{pmatrix}$ \\ \hline
6 & $\begin{pmatrix}
  & \lambda^6 & \\
  \lambda^6 & & \lambda^2 \\
  & \lambda^2 & \lambda^0
\end{pmatrix}$ &
$\begin{pmatrix}
  & \lambda^3 & \\
  & & \lambda^2 \\
  \lambda^1 & \lambda^0 & \lambda^0
\end{pmatrix}$ \\ \hline
\end{tabular}
\caption{The typical orders of matrix elements for six possible
mass textures. We have not explicitly included $O(1)$ coefficients,
which would be needed to precisely reproduce the experimental
data. Note that there are also additional solutions obtained (i) by
exchanging the second and third generations of $d_R$ (columns 
in $M_d$) and/or (ii) by identically relabeling generation indices
for $u_L$, $u_R$ and $d_L$.}
\label{oe}
\end{table}

Suitably choosing the $O(1)$ coefficients (i.e.\ Yukawa couplings) in
the textures listed in Table~\ref{oe}, we obtain numerical examples
for these six solutions (Table~\ref{predict}).
\begin{table}[thbp]
\renewcommand{\arraystretch}{1.25}
\centering
\begin{tabular}{c|cccccc} \hline\hline
& 1 & 2 & 3 & 4 & 5 & 6 \\ \hline
$m_u$ & ~~0.00179~~ & ~~0.00104~~ & ~~0.00179~~ & ~~0.00223~~ &
~~0.00179~~ & ~~0.000983~~ \\ \hline
$m_d$ & 0.00387 & 0.00470 & 0.00283 & 0.00321 & 0.00511 & 
0.00271 \\ \hline 
$m_s$ & 0.0562 & 0.0683 & 0.0637 & 0.0805 & 0.0566 & 0.0542 \\ \hline
$m_c$ & 0.613 & 0.611 & 0.652 & 0.633 & 0.601 & 0.621 \\ \hline
$m_b$ & 2.90 & 3.01 & 2.94 & 2.99 & 2.91 & 2.87 \\ \hline
$m_t$ & 178 & 177 & 175 & 176 & 179 & 171 \\ \hline
$|V_{us}|$ & 0.225 & 0.225 & 0.223 & 0.223 & 0.224 & 0.223 \\ \hline
$|V_{cb}|$ & 0.0433 & 0.0430 & 0.0433 & 0.0433 & 0.0428 & 
0.0434 \\ \hline
$|V_{ub}|$ & 0.00403 & 0.00414 & 0.00426 & 0.00429 & 0.00377 & 
0.00437 \\ \hline
$J_{\rm CP}\,/10^{-5}$ & 2.90 & 2.60 & 3.16 & 2.79 & 2.91 & 
2.83 \\ \hline
$\sin 2\phi_1/\beta$ & 0.709 & 0.692 & 0.762 & 0.735 & 0.694 & 
0.742 \\ \hline
$|(V_{dR})_{21}|$ & 0.0116 & 0.00990 & 0.144 & 0.167 & 0.285 & 
0.135 \\ \hline
$|(V_{dR})_{32}|$ & 0.681 & 0.682 & 0.643 & 0.642 & 0.656 & 
0.707 \\ \hline
$|(V_{dR})_{31}|$ & 0.0108 & 0.00922 & 0.125 & 0.147 & 0.0136 &
0.00918 \\ \hline
\end{tabular}
\caption{Numerical examples for the predictions of the texture
combinations given in Table~\ref{oe}. The mass eigenvalues are denoted
in GeV unit.}
\label{predict}
\end{table}


\end{document}